\documentclass[twocolumn,English,aps,pra,10pt,superscriptaddress,floatfix]{revtex4-2}
\usepackage{times}
\usepackage{graphicx}
\usepackage{amssymb}
\usepackage{bm}
\usepackage{amssymb,amsfonts,amsmath,amsbsy,bm,t1enc,latexsym}
\usepackage{float}
\usepackage[colorlinks=true,citecolor=blue,linkcolor=magenta]{hyperref}
\usepackage[markup=blue, authormarkupposition=left]{changes}
\usepackage{url}
\usepackage{xspace} 
\usepackage{siunitx}
\DeclareSIUnit \dbc {dBc}
\usepackage{soul}
\usepackage{changes}
\usepackage{cancel}
\usepackage{times}
\usepackage{soul}
\usepackage{amsmath}
\usepackage{hyperref}

\newcommand{\fref}[1]{Fig.~\ref{#1}}

\newcommand{\LN}[0]{$\mathrm{LiNbO}_3$} 
\newcommand{\LT}[0]{$\mathrm{LiTaO}_3$}

\begin{document}

\title{\Large{Stable Soliton Microcomb Generation in X-cut Lithium Tantalate via Thermal-Assisted Photorefractive Suppression}}







%
\author{Jiachen Cai}\thanks{These authors contributed equally.}
\affiliation{State Key Laboratory of Materials for Integrated Circuits, Shanghai Institute of Microsystem and Information Technology, Chinese Academy of Sciences, 865 Changning Road, Shanghai 200050, China.}
\affiliation{Center of Materials Science and Optoelectronics Engineering, University of Chinese Academy of Sciences,Beijing 100049, China}

\author{Shuai Wan}\thanks{These authors contributed equally.}
\affiliation{CAS Key Laboratory of Quantum Information, University of Science and Technology of China, 230026 Hefei, China.}
\affiliation{CAS Center for Excellence in Quantum Information and Quantum Physics, University of Science and Technology of China, 230026 Hefei, China.}

\author{Bowen Chen}\thanks{These authors contributed equally.}
\affiliation{State Key Laboratory of Materials for Integrated Circuits, Shanghai Institute of Microsystem and Information Technology, Chinese Academy of Sciences, 865 Changning Road, Shanghai 200050, China.}
\affiliation{Center of Materials Science and Optoelectronics Engineering, University of Chinese Academy of Sciences,Beijing 100049, China}

\author{Jin Li}\thanks{These authors contributed equally.}
\affiliation{CAS Key Laboratory of Quantum Information, University of Science and Technology of China, 230026 Hefei, China.}
\affiliation{CAS Center for Excellence in Quantum Information and Quantum Physics, University of Science and Technology of China, 230026 Hefei, China.}

\author{Xuqiang Wang}
\affiliation{State Key Laboratory of Materials for Integrated Circuits, Shanghai Institute of Microsystem and Information Technology, Chinese Academy of Sciences, 865 Changning Road, Shanghai 200050, China.}
\affiliation{Center of Materials Science and Optoelectronics Engineering, University of Chinese Academy of Sciences,Beijing 100049, China}

\author{Dongchen Sui}
\affiliation{State Key Laboratory of Materials for Integrated Circuits, Shanghai Institute of Microsystem and Information Technology, Chinese Academy of Sciences, 865 Changning Road, Shanghai 200050, China.}
\affiliation{Center of Materials Science and Optoelectronics Engineering, University of Chinese Academy of Sciences,Beijing 100049, China}

\author{Piyu Wang}
\affiliation{CAS Key Laboratory of Quantum Information, University of Science and Technology of China, 230026 Hefei, China.}
\affiliation{CAS Center for Excellence in Quantum Information and Quantum Physics, University of Science and Technology of China, 230026 Hefei, China.}

\author{Zhenyu Qu}
\affiliation{State Key Laboratory of Materials for Integrated Circuits, Shanghai Institute of Microsystem and Information Technology, Chinese Academy of Sciences, 865 Changning Road, Shanghai 200050, China.}
\affiliation{Center of Materials Science and Optoelectronics Engineering, University of Chinese Academy of Sciences,Beijing 100049, China}

\author{Xinjian Ke}
\affiliation{State Key Laboratory of Materials for Integrated Circuits, Shanghai Institute of Microsystem and Information Technology, Chinese Academy of Sciences, 865 Changning Road, Shanghai 200050, China.}
\affiliation{Center of Materials Science and Optoelectronics Engineering, University of Chinese Academy of Sciences,Beijing 100049, China}

\author{Yifan Zhu}
\affiliation{State Key Laboratory of Materials for Integrated Circuits, Shanghai Institute of Microsystem and Information Technology, Chinese Academy of Sciences, 865 Changning Road, Shanghai 200050, China.}
\affiliation{Center of Materials Science and Optoelectronics Engineering, University of Chinese Academy of Sciences,Beijing 100049, China}

\author{Yang Chen}
\affiliation{State Key Laboratory of Materials for Integrated Circuits, Shanghai Institute of Microsystem and Information Technology, Chinese Academy of Sciences, 865 Changning Road, Shanghai 200050, China.}
\affiliation{Center of Materials Science and Optoelectronics Engineering, University of Chinese Academy of Sciences,Beijing 100049, China}

\author{WenHui Xu}
\affiliation{State Key Laboratory of Materials for Integrated Circuits, Shanghai Institute of Microsystem and Information Technology, Chinese Academy of Sciences, 865 Changning Road, Shanghai 200050, China.}
\affiliation{Center of Materials Science and Optoelectronics Engineering, University of Chinese Academy of Sciences,Beijing 100049, China}

\author{Ailun Yi}
\affiliation{State Key Laboratory of Materials for Integrated Circuits, Shanghai Institute of Microsystem and Information Technology, Chinese Academy of Sciences, 865 Changning Road, Shanghai 200050, China.}
\affiliation{Center of Materials Science and Optoelectronics Engineering, University of Chinese Academy of Sciences,Beijing 100049, China}

\author{Jiaxiang Zhang}
\affiliation{State Key Laboratory of Materials for Integrated Circuits, Shanghai Institute of Microsystem and Information Technology, Chinese Academy of Sciences, 865 Changning Road, Shanghai 200050, China.}
\affiliation{Center of Materials Science and Optoelectronics Engineering, University of Chinese Academy of Sciences,Beijing 100049, China}

\author{Chengli Wang}
\affiliation{State Key Laboratory of Materials for Integrated Circuits, Shanghai Institute of Microsystem and Information Technology, Chinese Academy of Sciences, 865 Changning Road, Shanghai 200050, China.}
\affiliation{Center of Materials Science and Optoelectronics Engineering, University of Chinese Academy of Sciences,Beijing 100049, China}

\author{Chun-Hua Dong}
\email[]{chunhua@ustc.edu.cn}
\affiliation{CAS Key Laboratory of Quantum Information, University of Science and Technology of China, 230026 Hefei, China.}
\affiliation{CAS Center for Excellence in Quantum Information and Quantum Physics, University of Science and Technology of China, 230026 Hefei, China.}

\author{Xin Ou}
\email[]{ouxin@mail.sim.ac.cn}
\affiliation{State Key Laboratory of Materials for Integrated Circuits, Shanghai Institute of Microsystem and Information Technology, Chinese Academy of Sciences, 865 Changning Road, Shanghai 200050, China.}
\affiliation{Center of Materials Science and Optoelectronics Engineering, University of Chinese Academy of Sciences,Beijing 100049, China}

\maketitle

\noindent
{\large\textbf{Abstract}}

\textbf{Chip-based soliton frequency microcombs combine compact size, broad bandwidth, and high coherence, presenting a promising solution for integrated optical telecommunications, precision sensing, and spectroscopy. Recent progress in ferroelectric thin films, particularly thin-film Lithium niobate (\LN) and thin-film Lithium tantalate (\LT), has significantly advanced electro-optic (EO) modulation and soliton microcombs generation, leveraging their strong third-order nonlinearity and high Pockels coefficients. However, achieving soliton frequency combs in X-cut ferroelectric materials remains challenging due to the competing effects of thermo-optic and photorefractive phenomena. These issues hinder the simultaneous realization of soliton generation and high-speed EO modulation. Here, following the thermal-regulated carrier behaviour and auxiliary-laser-assisted approach, we propose a convenient mechanism to suppress both photorefractive and thermal dragging effect at once, and implement a facile method for soliton formation and its long-term stabilization in integrated X-cut \LT~microresonators for the first time. The resulting mode-locked states exhibit robust stability against perturbations, enabling new pathways for fully integrated photonic circuits that combine Kerr nonlinearity with high-speed EO functionality.}\\


\noindent
{\large\textbf{Introduction}}

Harnessing the enhanced light-matter interaction inside the micro-nano optical cavities, frequency comb have triggered the development of multi-channel source generation via the nonlinearity-driven four wave mixing (FWM) \cite{del2007optical,herr2012universal}. By balancing group velocity dispersion (GVD) and inherent Kerr nonlinearity, dissipative Kerr soliton (DKS) -- mode-locked pulses by high power pumping -- offers low noise, high coherence and static comb spectra \cite{herr2016dissipative,herr2014temporal}. These features have advanced applications in optical telecommunication \cite{shu2022microcomb,zhang2024high}, microwave engineering \cite{liu2020photonic,zhao2024all} and LiDAR \cite{trocha2018ultrafast,chen2023breaking}. Rapidly deployment of large-volume manufacturing has made chip-integrated frequency combs accessible across various thin-film materials, including silica \cite{yang2017stokes,lee2017towards}, $\mathrm{Si_3N_4}$ \cite{brasch2016photonic,shen2020integrated}, AlGaAs \cite{chang2020ultra,wu2022soliton}, \LN~\cite{wang2019monolithic,song2024octave}, and \LT~\cite{wang2024lithium}. More importantly, multi-layer heterogeneous integration and efficient planar microfabrication enable the integration of multiple optical devices onto a single chip, revolutionizing the architectures of photonic integrated circuits (PICs) and frequency microcombs.
Among these novel materials, \LN~and \LT, categorized as ferroelectric crystals, emerge as promising candidates for versatile photonic platforms due to their high-linear EO modulation, plentiful non-linear optical effects and ultralow propagation loss \cite{wang2018integrated,boes2023lithium,qi2020integrated,wang2024ultrabroadband,wang2024lithium,yu2024tunable}.

\begin{figure*}[htbp]
	\centering
	\includegraphics[scale = 1]{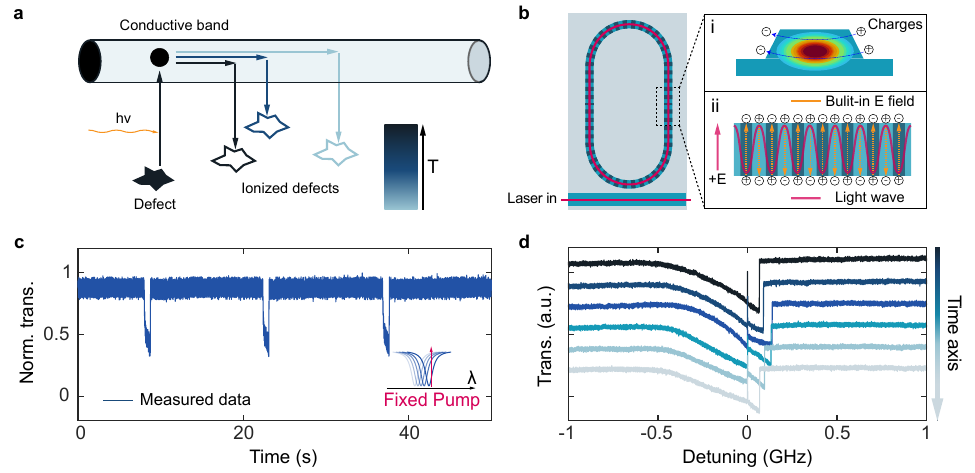}
	\caption{\textbf{Photorefractive effect results in the uncertainty of soliton comb generation in X-cut \LT.} 
	(a) Energy-level diagram illustrates the photorefractive effect inside the \LT~material. Higher temperature contributes to a higher carrier conductivity and in turn suppresses the lifetime of carriers activated by photorefractive effect. (b) Principle diagram of photorefractive-induced EO modulation (i) and cavity-grating construction created by photovoltaic electric field (ii). 
    (c) Measured transmission with a pump fixed in the red-detuned side of the resonance of an air-cladded x-cut \LT~ racetrack microresonator. The expected hysteresis can be attributed to the (i) situation in (b). (d) Measured scanning spectra near a resonance, with a scan rate up to 65 GHz/s and an on-chip laser power of near 32 mW. Its temporal evolution proves that the photorefractive optical grating evolving over time.
	}
	\label{fig1}
\end{figure*}

Given the excellent nonlinear features within the ferroelectrics, it naturally raises up the conceptional combination of a high-speed EO modulator and a broadband Kerr comb source to establish a fully stabilized microcombs. To date, Kerr soliton microcomb has been demonstrated in Z-cut \LN~\cite{gong2019soliton,gong2020near,wang2024octave,song2024octave}, particularly with bi-started characteristic \cite{he2019self} and mode-locked state self-stability \cite{wan2024photorefraction}. However, this thin-film cut type physically limits access to the maximum Pockels coefficient ($r_{33}$), impeding the integration of high-performance EO modulators with Kerr combs. Although seeding Kerr combs is feasible in X-cut \LT~\cite{wang2024lithium}~or \LN~\cite{yu2020raman}, achieving soliton frequency combs with long equilibration times remains a significant challenge. This is due to the slow blue shift caused by photorefractive effect and the rapid red shift caused by the thermo-optic (TO) effect, which together create a dynamic disequilibrium that hampers the stabilization of mode-locked states. 
Only recently the rapid single sideband sweep was investigated for the initial addressing of soliton states in ferroelectric cavities \cite{wang2024lithium}, but the triggered soliton state trends to disappear within a few seconds, since the detuning condition on which the soliton generation depends will vary with the refractive index change on a time scale of seconds. Although this resonance excursion can be tracked by the power-kicking \cite{brasch2016photonic} or Pound–Drever–Hall (PDH) laser locking techniques \cite{stone2018thermal}, the electronic devices are sophisticated and not suitable for PICs with miniaturization requirements. Thus, for the realization of Kerr soliton microcomb with long equilibration time in X-cut ferroelectrics, it is essential to eliminate the complex photorefractive disruptions affecting micrometre-sized waveguides. This approach could further simplify the search for Kerr soliton states, and support the development of low-voltage-driven modulators in a monolithic way. While techniques such as annealing and cladding removal have been employed to mitigate the PR effect \cite{xu2021mitigating,xu2021bidirectional}, their effectiveness in stabilizing soliton microcomb under high optical power conditions has not been sufficiently studied yet.


In this paper, we realize the photorefractive-free resonators by simply heating up the \LT-based microresonator. Without the need for an additional electronic feedback loop to dynamically stabilize the laser-resonance offset, we demonstrate Kerr soliton generation using the dual-suppressed strategy, which involves mitigating photorefractive effects through thermally controlled carrier dynamics and minimizing TO effects via an auxiliary-assisted approach. This comb generation has proven to be highly reliable, with only manually or piezo-module tuning of the optical instruments. Notably, the deterministic generation and prolonged lifetime of mode-locked states represent significant advantages of this technique, overcoming the limitations posed by time-dependent photorefractive phase modulation and the huge thermal dragging response in X-cut ferroelectric photonic platforms. 
~\\
\\
{\large\textbf{Results}}

\begin{figure*}[htp]
	\centering
	\includegraphics[scale = 1]{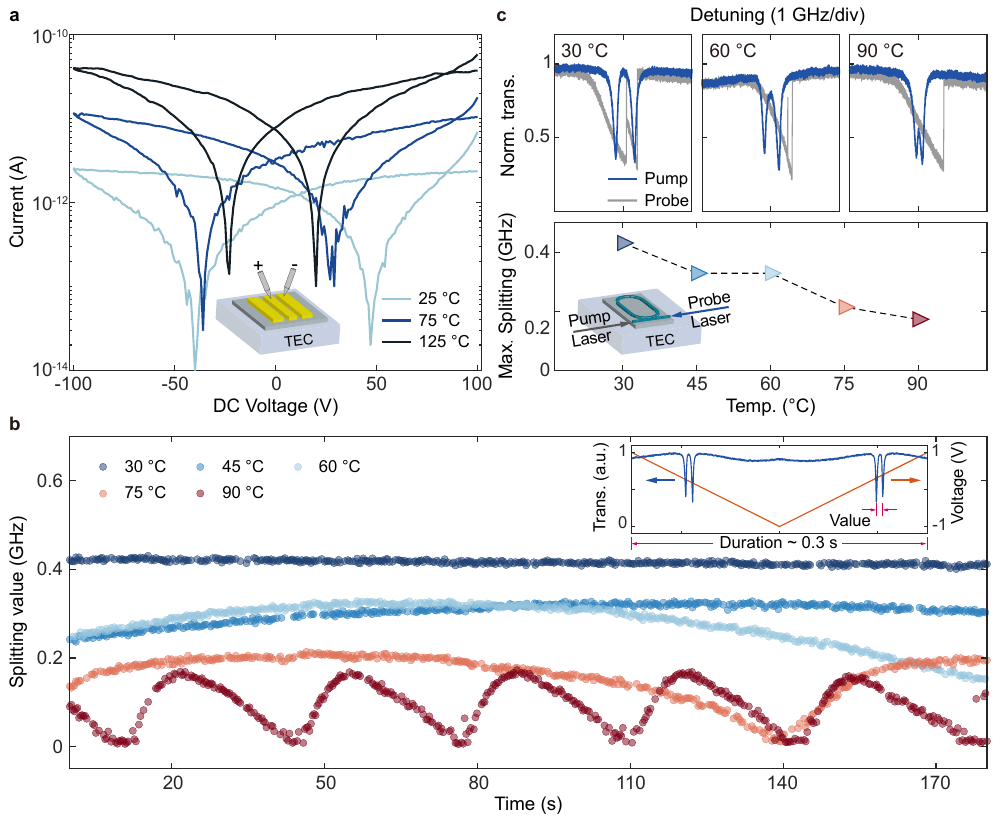}
	\caption{\textbf{Suppressed photorefractive effect by means of thermoelectric cooler (TEC)-controlled heating.} 
     (a) I-V curve with maximum $\pm$100 V bias voltage in CPW with 6 $\mu$m gap (the inset image). Higher conductivity and less hysteresis level are observed with higher temperature. (b) Temporal variation of the splitting value at different temperatures with three-minute sweeping characterization. Missing sampling points, which are largely obscured by the background noise, do not impact the following analysis of the hysteresis phenomenon. The inset represents a splitting value extracted from a piezo-actuated probe laser (blue line) controlled by an electrical triangle wave (orange line). (c) Upper panel: close-up transmission of pump laser (grey) and probe laser (blue) under different temperature conditions. Lower panel: statistic correlation of maximum mode splitting with temperature from (b), manifesting a nonlinear decline in the splitting rendered by photorefractive grating. 
	}
	\label{fig2}
\end{figure*}

We start with a simple band model for illustrating the photorefractive mechanism in the ferroelectric material, as displayed in \fref{fig1}(a). Defect levels between the valance band and the conduction band are ubiquitous in congruent ferroelectric materials and are thought to originate from vacancies and antisite defects inside the crystal lattice \cite{imbrock2003photorefractive}, as well as defects introduced during bonding and etching processes. These defects often match the photon energy, allowing them to interact with the incident light. Depending on the laser intensity and photon frequency, photoexcited carriers are promoted to the conduction band, where they transport before either recombining with other carriers or returning to the defect levels. This electron process can produce a refractive index change in ferroelectric materials, known as the photorefractive effect \cite{gunter1988photorefractive}. In waveguide components, the electric fields of confined optical modes direct the movement of these photo-excited carriers. For instance, in transverse-electric (TE) modes, the horizontal potential difference created by the optical field leads to a photovoltaic electric field and hence to the Pockels modulation, as carriers accumulate along the direction opposite to the electric field component of the light mode (\fref{fig1}(b)(i)).
In comparison with the transient light-matter interaction, the electronic reaction has a larger relaxation time, which stands for a longer period of the carrier migration and results in the emergence of a slowly developing space-charge field. Based on this effect, Z-cut soliton can be accessed via manually tuning the laser frequency into the cavity resonance \cite{gong2019soliton,wan2024photorefraction}. However, the circumstances of in-plane optical mode pumping in X-cut \LN~or \LT~resonators are markedly distinct from the Z-cut variant, primarily owing to its larger thermo-optic (TO) coefficient \cite{zhang2023fundamental}. Although the rapid TO shift locks the pump laser within the thermal bistability regime \cite{herr2014temporal}, it also narrows the soliton existence step and renders the soliton seeding operations ineffective \cite{li2017stably}. Therefore, the opposite phase shift produced by the photorefractive effect might over-compensate the thermal resonance drift to a certain extent, and the distinction between above two reactions is also manifested in the timescale, whereby the postponed photorefractive behaviour exacerbates the drawback of locking the laser-resonance offset to an accessible DKS state. \fref{fig1}(c) reflects that the pump-resonance offset jitters when the pump laser is fixed in the red-detuned resonance region. A dramatic blue shift happens slowly once tuning the laser into the cavity resonance dip from the red-detuned side, and vice versa. At the same time, the swift TO shift towards the long wavelength regime would result in the pump exceeding the cavity resonance, which has appeared in previous works \cite{sun2017nonlinear,wang2019monolithic,cai2024high}, betraying the competitive relationship between two effects and its irresistible prohibition to emerging mode-locked comb state. In the optical microresonator, the standing wave characteristic shapes the spatial distribution of the built-in electric field (\fref{fig1}(b)(ii)), creating the refractive-index-modulated pattern in analog to the grating with the frequency selective feature \cite{hou2024subwavelength}. With high-power pump frequency tuning over the resonance, the formed mode splitting exhibits an apparent evolution (\fref{fig1}(d)) that can be also accounted for the enhanced charge-discharge-type electric field originating from the production and the neutralization of carriers at room temperature \cite{liu2021resonant,xu2021photorefraction}. This introduces the instability into the nonlinear frequency conversion process, such as the deviation in local integrated dispersion and the reduction in the extinction ratio of the pumped resonance \cite{wang2024coupling}. As a result, a more complicated pumping mechanism is required to enable the soliton formation.

Heating is a plausible technique for degenerating photorefractive effect in the confined optical waveguide, which has been employed for nonlinear harmonic generation \cite{lu2020toward,lu2021ultralow,cheng2024efficient} and EO modulation \cite{celik2024roles}. According to the classical photorefractive principle \cite{gunter1988photorefractive}, the carrier lifetime $\tau$ is inversely proportional to the ion conductivity $\sigma$. Based on the Arrhenius-type relationship \cite{ruprecht2012low,lucas2022high}, $\sigma(T) \propto e^{-{\frac{1}{T}}}$, it is prone to enhance the ion conductivity and subsequently reduce the lifetime $\tau$ via elevated temperature operation. This behaviour is verified by the I-V measurement of a coplanar waveguide (CPW) electrode on thin film. As shown in \fref{fig2}(a), an exponential growth in current is observed, consistent with the conductivity enhancement due to the heating. Furthermore, the hysteresis observed during voltage switching decreases as the temperature rises from 25 $^{\circ}\mathrm{C}$ to 125 $^{\circ}\mathrm{C}$, reflecting the accelerated carrier dynamics at higher temperatures. 
For a more thorough illustration, we set up a modulator/demodulator system (Supplementary Note S1) to conduct a photorefractive grating measurement based on our X-cut \LT~microresonators by planer manufacturing, which is described in detail in the Methods section. The grating-induced separation between clockwise (CW) and counter-clockwise (CCW) modes, as inferred from the probe transmission in the inset of \fref{fig2}(b), is defined as the splitting value. To measure the temporal change of the pumped optical resonance, the mode splittings periodically scanned by a low power laser is recorded at a sampling interval of 0.3 seconds for 3 minutes. With the increase in temperature from 30 $^{\circ}\mathrm{C}$ to 90 $^{\circ}\mathrm{C}$, the mode-splitting curves display a numerical decline and a shorter period of oscillation at higher temperature (\fref{fig2}(b)). It could be deduced that the attenuated grating behaviour is a consequence of a reduced existence time of the heated carriers, and that the diffusion and recombination of accelerated charge carriers are responsible for a more obvious charge-discharge cycle for photovoltaic electric field. Correspondingly, \fref{fig2}(c) plots the photorefractive grating appearance (upper panel) and the maximum resonance splitting degrees versus temperature (lower panel), enabling access to on-demand manipulating the photorefractive effect inside a high-Q microcavity under the temperature-assisted operation.

\begin{figure*}[htbp]
	\centering
	\includegraphics[scale = 1]{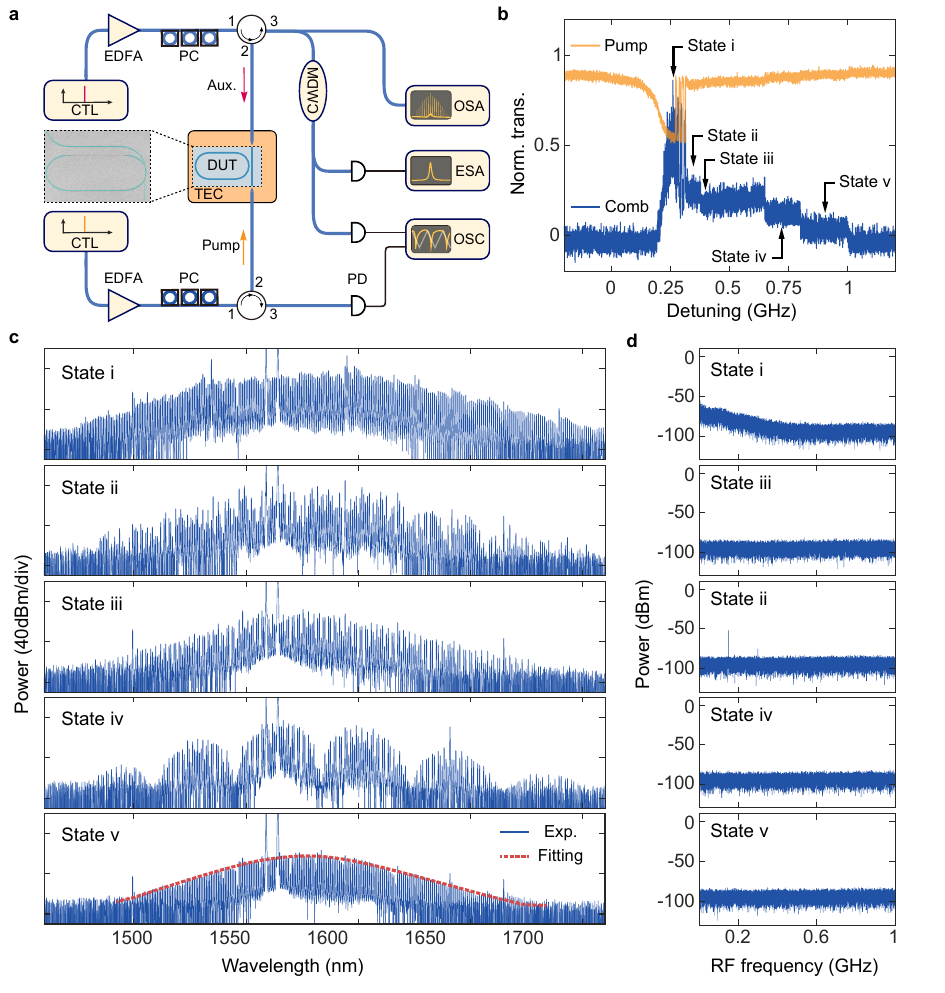}
	\caption{\textbf{Soliton frequency comb generation in the TFLT-based racetrack microresonator at 230 $^{\circ}\mathrm{C}$.}
        (a) The experimental setup via dual laser pumping. DUT, device under test, CTL, continuously tunable laser, EDFA, erbium-doped fiber amplifier, PC, polarization controller, PD , photodetector, CWDM, coarse wavelength division multiplexer, ESA, electrical spectrum analyzer, OSC, oscilloscope, OSA, optical spectrum analyzer. The inset is the scanning electron microscope (SEM) of racetrack resonator for soliton generation. (b) The transmission spectrum of the pump (yellow) and the comb power (dark blue). With the help of auxiliary laser, the pump laser frequency is scanned from the blue-detuned regime to the red-detuned regime of the pump resonance to acquire the soliton steps in the red-detuned region. (c) Spectra snapshots recorded at five different stages marked in (b), corresponding to MI comb (State i), multi-soliton (State ii), breather soliton (State iii), two-soliton (State iv) and single soliton (State v). (d) RF amplitude noise spectra corresponding to the five different states.
	}
	\label{fig3}
\end{figure*}

To generate DKS soliton relying on the foregoing finding, a dual-suppressed strategy is performed using our experimental setup, as shown schematically in \fref{fig3}(a). First, to counteract the unwanted photorefractive modulation while pumping the cavity with more than one hundred milliwatts order of magnitude, stronger heating operation via TEC is considered in the experiment to meet the demand of no additional resonances drift. Second, as for the cavity temperature stabilization under huge TO effect, the thermal dragging dynamics that exists in soliton formation is bypassed using the dual-laser-driven method \cite{zhou2019soliton}. With a total insertion loss of approximately 10 dB (5dB per facet from fiber-to-chip coupling via lens fiber and negligible loss from optical chip), two high-power sources are employed to inject into the cavity from two opposite direction after fiber optical system including power amplifying, polarization controlling and edge coupling. Following a conventional tuning trace from blue regime to red regime, a 1559 nm transverse-magnetic (TM) polarized optical source with 185 mW on-chip power, serving as the auxiliary laser, would level off to the resonance dip in advance. In the meantime, the TEC-controlled temperature of the chip holder should be adjusted finely till the time-varying resonance excursion vanishes and thermal resonance triangle dominates. The final settled temperature is set to 230 $^{\circ}\mathrm{C}$ to fully compensate the carrier relaxation time in photorefractive effect. As such, the auxiliary laser can be adjusted close to the blue-detuned regime of the TM mode resonance. The low quality factor and overwhelming anomalous dispersion of TM mode group (See Supplementary Note S2) raise the microcomb generation threshold, which is preferable to suppress the unfavorable beatnote noise arising from different microcombs generated by two high-power sources \cite{lambert2023microresonator}. To get soliton microcomb seeding started, a 1564 nm tunable external-cavity diode laser, with 97 mW on-chip power and TE polarization and 0.75 GHz/ms piezo-controlled laser sweeping, is required to tune near the high-Q TE mode resonance. Throughout the soliton formation process, the abrupt alteration in pump power, which corresponds to the transition of microcomb states, creates a passive tuning of auxiliary laser-resonance offset due to the TO effect. This equivalent change in intracavity auxiliary power in turn maintains the total intracavity power to guarantee the thermal balance of all cavity resonances.

\fref{fig3}(b) shows the generated comb power and optical transmission, stable comb power profile betrays a simultaneous suppression of the photorefractive effect and the thermal effect. When the pump laser is biased from the short wavelength regime to the long wavelength regime with respect to the pumped resonance, the cavity power exhibits oscillating optical intensity followed by multistage flattened steps. This signal delineates a specific progression from the modulation instability (MI) state to the low-noise mode-locked states. By setting the pump laser rigidly in the effective blue-detuning steps, differnet types of temporal dissipate soltion waveforms can be witnessed in \fref{fig3}(c), including MI comb (state i),  breather soliton comb (state ii), multi-soliton comb (state iii), two-soliton comb (state iv) and single soliton comb (state v), with RF noise spectra confirming their coherence (\fref{fig3}(d)). Note that a perfect soliton crystal state with 13-FSR spacing comb lines is observed on the X-cut thin film \LT~for the first time (See Supplementary Note S3). The available extent of the microcomb spectra can be optimized through the elaborate dispersion design and higher power pumping, so as to establish the octave spanning microcomb catching up to the state-of-the-art performance in Z-cut ferroelectric platform \cite{song2024octave}. As a proof of concept in our work, this method similarly provides an universal method for accessing DKS comb on X-cut thin film \LN~photonic platform \cite{yu2020raman}.

\begin{figure*}[htp]
	\centering
	\includegraphics[scale = 1]{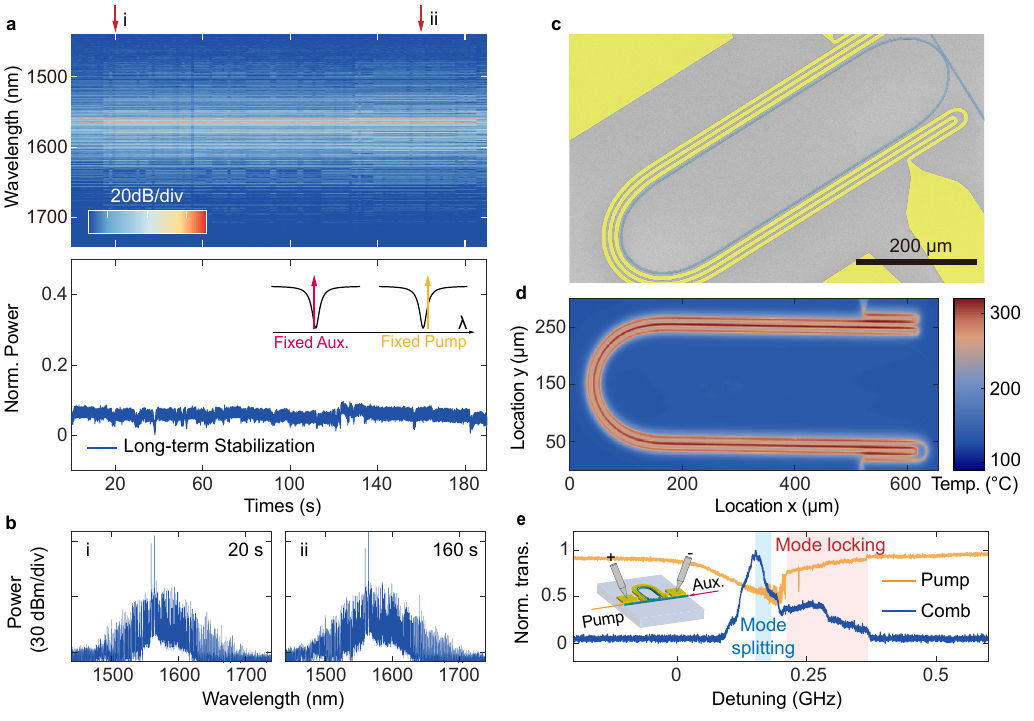}
	\caption{\textbf{Mode-locked frequency comb generation with its long-live properties and device architecture for monolithic mode-locked state formation.}
        (a) Measured wavelength-time mapping of the mode-locked state and normalized comb power versus time with fixed red-side pump laser and fixed blue-side auxiliary laser. The displayed jitters imply the transitions between different multi-soliton states. (b) Silced spectral pictures at the time of 20s and 160s, which verifies the long-term stability of mode-locked state via thermal-assisted technique. (c) SEM pictures of \LT~racetrack resonator with integrated spiral heater. (d) Temperature distribution of on-chip heater applied under 80 V DC voltage. (e) Transmission spectrum of the pump (yellow) and the comb power (dark blue), the typical stepwise comb power in the red-shaded region shows typical mode-locked state formation.   
	}
	\label{fig4}
\end{figure*}

Importantly, to showcase the potential of our dual-suppressed technique, we test the long-term performance of the multi-soliton state via fixing a TE-polarized pump laser (1564.240 nm) and a TM-polarized auxiliary laser (1559.147 nm) engaged by an elevated temperature to heat up the cavity. The absence of the resonance drift caused by the photorefractive-induced charge field enables the mode-locked comb existence exceeding 3 minutes, as depicted in the time-wavelength map in \fref{fig4}(a). In \fref{fig4}(b), sliced spectral frames at the fixation time of 20 s and 160 s provide further details on that the representative multi-soliton-shaped envelope is well preserved over long equilibration periods in our experiment. Note that the power fluctuations of the depicted comb intensity are ascribed to the temperature instability of the TEC and the variations of fiber-to-chip coupling caused by the thermal airflow. These disruptions do not impair the mode-locked nature of our generated microcomb, providing compelling advantages for photonic applications requiring stable comb sources.

Apart form the redundancies of complex electronic devices to access and stabilize the mode-locked soliton, the heat-assisted approach faces limitations due to the large device volume of TEC-controlled equipment and unavoidable heating-air perturbation affecting the edge coupling system. To address these issues, we fabricated the integrated Platinum (Pt) heater (See Methods) with a spiral channel structure designed to form and localize the heating field, enabling a fully integrated soliton generation prototype (\fref{fig4}(c)). This greatly reducing the footprint of the soliton generator. By applying an 80 V DC voltage to an electrode with approximately 2162 ohms resistance, we can get the heat distribution map of the etched surface using a temperature measurement microscope systems (QFI/InfraScope-TM-HS).
The localized thermal domain near the waveguide reached temperatures exceeding 200 $^{\circ}\mathrm{C}$ (\fref{fig4}(d)), confirming the effectiveness of the on-chip heater. Based on the same dual-laser-driven configuration, we reproduced the microcomb generation via sweeping the pump laser. As inferred from the normalized comb power (\fref{fig4}(e)), the step-like comb power confirms the triggering of different mode-locked states similar to the aforementioned results. However, the use of the localized thermal field reduces device volume and minimizes coupling issues, demonstrating the advantages of the integrated approach.
~\\
\\
{\large\textbf{Discussion}}


In summary, the intracavity photorefractive effect in the X-cut \LT~thin film can be readily restrained at high temperature, as thermally improving carrier conductivity breaks the non-static balance between internal charge field and intracavity electromagnetic waves. Via dual-laser-driven scheme, we obtain various soliton states in a low-loss \LT~cavity, including soliton crystals and single soliton. Our results about the robustness of low noise state with considerable equilibration lifetime in experiments and all-on-chip demonstration are core to applications involving with fast reconfigurable soliton microcomb \cite{he2023high,ling2024electrically} based on X-cut ferroelectric platform.

\begin{figure}[htp]
	\centering
	\includegraphics[scale = 1]{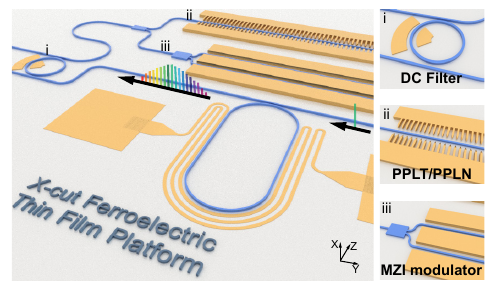}
	\caption{Schematic outlook of fully-integrated optoelectronic chip based on ferroelectric materials, compatible with coherent Kerr comb source, DC-stable ring filter (i), periodically-poled components (ii) and high-speed EO modulation (iii). 
	}
	\label{fig5}
\end{figure}

Nevertheless, there are several issues requiring further attention. The Pt-crafted heater, like other metals such as titanium and chromium, cannot endure prolonged exposure to high voltages, which may lead to severe electrical breakdown. Similarly, the air-exposed Joule heating is unattainable with long-term reliable operation due to the penetrated air attempts to oxidize the metal at elevated temperatures. Moreover, the experimental results for DKS steps are not ideal, shown in the blue shaded regions in \fref{fig4}(e). This can be attributed to residual photorefractive effects and coupling fluctuations, leading to a mode splitting and the ongoing resonance instability. To address the existing problems, some solutions relevant to this work are promising. For example, the thermal management by advanced device structure can be utilized to optimize heat distribution \cite{bar2021chip} and reduce the heating impact on the coupling stability. Modified microheater microstructures, especially in the realm of microelectronics \cite{yu2023chip,jin2024chip} and micro-electromechanical system (MEMs) \cite{salvia2009real,popa2021highly}, might enable a loop in a more reliable thermal field control. Tricky ways, such as rising soliton steps in larger FSR resonators \cite{wu2022soliton} to weaken the theraml response, can reduce the need of auxiliary laser and further miniaturize the chip footprint. 

Broadly speaking, we believe that the thermal-assisted approach has the potential of enriching the current microcomb generation for all-optical \LN~or \LT~integrated circuit. As illustrated in \fref{fig5}, this platform, highly compatible with high-speed modulation \cite{wang2018integrated,wang2024ultrabroadband} and horizontal periodically poled waveguide offering high nonlinear conversion efficiency \cite{hwang2023tunable,shi2024efficient,chen2023periodic}, positioning it as pivotal for next-generation optical applications. 
A notable application of this technique lies in coherent optical communications \cite{li2009recent,marin2017microresonator,liu2024parallel}. As well as empowering hyperscale data transmission network with high-coherence parallelization and high-speed reconfiguration, the monolithic realization of stable DKS formation without electrical feedback and $r_{33}$-based EO modulation can also avoid the device complexity of optoelectronic systems. Combined with the periodically poled components, the versatility of this X-cut design also facilitates the frequency doubling applications from visible to infrared band, such as selective harmonic generation \cite{li2024advancing}, broadband synthetic frequency lattice \cite{javid2023chip} and efficient optical parametric amplifier \cite{stokowski2024integrated} based on coherent multi-channel sources. 
~\\
\\
{\large\textbf{Methods}}
\\
\textbf{Device fabrication}
The device fabrication involves two processes: the manufacturing of thin film \LT~ and the planer device fabrication. Via ion cutting and wafer bonding (commonly termed smart-cut technique), 600-nm-thickness \LT~thin film is sliced from an optical-grade \LT~bulk wafer to the thin-film-on-insulator structure, including 600 nm LT thin film, 4.7 $\mu$m oxide layer and 500 $\mu$m silicon substrate, subsequently followed with dicing into small chips with 1 cm $\times$ 1.2 cm for a more flexible device demonstration. With regard to the device fabrication, the micro-nano patterns are transferred onto the diced chips using e-beam lithography (Elionix ELS-BODEN 125) with hard mask (Ma-N 2405) and physical dry etching process with argon ions (Leuven, HAASRODE-I200). Next, an alkaline wet etching approach is employed to remove the sidewall redeposition as well as the etched resist. The well-defined microheater, comprising 150 nm Pt and 10 nm Ti, is fabricated through a dual-layer lift-off process, which employs maskless lithography (Heidelberg DWL 66+) and e-beam metal evaporation system.
Our under-test \LT~microresonator (inset of \fref{fig3}(a)) has an average intrinsic Q value of 3.6 million and 100 GHz FSR, with 400 nm etched depth and 200 nm slab. In order to generate robust bright soliton state, the balance between material nonlinearity and group velocity dispersion can be obtained by anomolous integrated dispersion (See Supplementary Note S2), with a multimode waveguide width of 2 $\mu$m. Alternatively, the adopted racetrack device is intended to be orientation selective so that the Raman scattering in microresonators is inhibited to ensure the maximum conversion efficiency of Kerr nonlinear process. 
\\
\textbf{Characterization of Q-factor}
For the Q-factor characterization, we use the Toptica DLC CTL 1550 to deliver the tunable continuous-wave laser into the racetrack reasonator with specific polarization, which is realized by variable optical attenuator (VOA), polarization controller (PC) and tapered fiber coupling. To extract the actual group dispersion from the laser sweeping measurement with a nonlinear piezo-driven property, a customized fiber-based MZI interferometer, with 50 m delay line, is used to calibrate the definite location of resonances. Measured intrinsic loss rate distribution is indicated in Supplementary Note S2, where the average $\kappa_0{/}2\pi$ is approximately 53 MHz revealing a mean intrinsic Q-factor of near $3.6 \times 10^6$.
\newline


\section*{Acknowledgments}
The sample fabrication in this work is supported by Shanghai Institute of Microsystem and Information Technology (SIMIT) material-device process and characterization platform, ShanghaiTech Material and Device Lab (SMDL), JFS laboratory and the USTC Center for Micro and Nanoscale Research and Fabrication.
This work has been supported by the National Key Research and Development Program of China (2022YFA1404601), the National Natural Science Foundation of China (62293520, 62293521, 12074400, 62205363, 12104442, 12404446, 12293052), Shanghai Science and Technology Innovation Action Plan Program (20JC1416200, 22JC1403300), CAS Project for Young Scientists in Basic Research (Grant No. YSBR-69), Anhui Provincial Natural Science Foundation (Grant No. 2408085QA010), the China Postdoctoral Science Foundation (Grant No. 2024M753078), the Postdoctoral Fellowship Program of CPSF (GZC20232560).
\newline

\section*{Author contributions} 
X.K., C.Y. and X.O. fabricated the \LT~wafers.
J.C., C.W. and S.W. designed the devices.
J.C., B.C., X.W. and Y.Z. fabricated the devices.
J.L., J.C.,  P.W., B.C., D.S, W.X. and Z.Q. carried out the measurements of photorefractive-induced mode splitting, soliton generation and the I-V curve.
J.C. and J.L. analyzed the data.
J.C., C.W. and S.W prepared the figures and wrote the manuscripts with contributions from all authors.
C.W., C.D. and X.O. supervised the project.

\section*{Competing interests}
The authors declare no competing financial interests.

\section*{Data Availability Statement} The code and data underlying the results presented in this work are not publicly available but can be obtained from the authors upon reasonable request.

\bibliography{refs}	
\bibliographystyle{naturemag}

\end{document}


\title{{\Large{Supplementary Information:}} \\ Stable Soliton Microcomb Generation in X-cut Lithium Tantalate via Thermal-Assisted Photorefractive Suppression}

\author{Jiachen Cai}\thanks{These authors contributed equally.}
\affiliation{State Key Laboratory of Materials for Integrated Circuits, Shanghai Institute of Microsystem and Information Technology, Chinese Academy of Sciences, 865 Changning Road, Shanghai 200050, China.}
\affiliation{Center of Materials Science and Optoelectronics Engineering, University of Chinese Academy of Sciences,Beijing 100049, China}

\author{Shuai Wan}\thanks{These authors contributed equally.}
\affiliation{CAS Key Laboratory of Quantum Information, University of Science and Technology of China, 230026 Hefei, China.}
\affiliation{CAS Center for Excellence in Quantum Information and Quantum Physics, University of Science and Technology of China, 230026 Hefei, China.}

\author{Bowen Chen}\thanks{These authors contributed equally.}
\affiliation{State Key Laboratory of Materials for Integrated Circuits, Shanghai Institute of Microsystem and Information Technology, Chinese Academy of Sciences, 865 Changning Road, Shanghai 200050, China.}
\affiliation{Center of Materials Science and Optoelectronics Engineering, University of Chinese Academy of Sciences,Beijing 100049, China}

\author{Jin Li}\thanks{These authors contributed equally.}
\affiliation{CAS Key Laboratory of Quantum Information, University of Science and Technology of China, 230026 Hefei, China.}
\affiliation{CAS Center for Excellence in Quantum Information and Quantum Physics, University of Science and Technology of China, 230026 Hefei, China.}

\author{Xuqiang Wang}
\affiliation{State Key Laboratory of Materials for Integrated Circuits, Shanghai Institute of Microsystem and Information Technology, Chinese Academy of Sciences, 865 Changning Road, Shanghai 200050, China.}
\affiliation{Center of Materials Science and Optoelectronics Engineering, University of Chinese Academy of Sciences,Beijing 100049, China}

\author{Dongchen Sui}
\affiliation{State Key Laboratory of Materials for Integrated Circuits, Shanghai Institute of Microsystem and Information Technology, Chinese Academy of Sciences, 865 Changning Road, Shanghai 200050, China.}
\affiliation{Center of Materials Science and Optoelectronics Engineering, University of Chinese Academy of Sciences,Beijing 100049, China}

\author{Piyu Wang}
\affiliation{CAS Key Laboratory of Quantum Information, University of Science and Technology of China, 230026 Hefei, China.}
\affiliation{CAS Center for Excellence in Quantum Information and Quantum Physics, University of Science and Technology of China, 230026 Hefei, China.}

\author{Zhenyu Qu}
\affiliation{State Key Laboratory of Materials for Integrated Circuits, Shanghai Institute of Microsystem and Information Technology, Chinese Academy of Sciences, 865 Changning Road, Shanghai 200050, China.}
\affiliation{Center of Materials Science and Optoelectronics Engineering, University of Chinese Academy of Sciences,Beijing 100049, China}

\author{Xinjian Ke}
\affiliation{State Key Laboratory of Materials for Integrated Circuits, Shanghai Institute of Microsystem and Information Technology, Chinese Academy of Sciences, 865 Changning Road, Shanghai 200050, China.}
\affiliation{Center of Materials Science and Optoelectronics Engineering, University of Chinese Academy of Sciences,Beijing 100049, China}

\author{Yifan Zhu}
\affiliation{State Key Laboratory of Materials for Integrated Circuits, Shanghai Institute of Microsystem and Information Technology, Chinese Academy of Sciences, 865 Changning Road, Shanghai 200050, China.}
\affiliation{Center of Materials Science and Optoelectronics Engineering, University of Chinese Academy of Sciences,Beijing 100049, China}

\author{Yang Chen}
\affiliation{State Key Laboratory of Materials for Integrated Circuits, Shanghai Institute of Microsystem and Information Technology, Chinese Academy of Sciences, 865 Changning Road, Shanghai 200050, China.}
\affiliation{Center of Materials Science and Optoelectronics Engineering, University of Chinese Academy of Sciences,Beijing 100049, China}

\author{WenHui Xu}
\affiliation{State Key Laboratory of Materials for Integrated Circuits, Shanghai Institute of Microsystem and Information Technology, Chinese Academy of Sciences, 865 Changning Road, Shanghai 200050, China.}
\affiliation{Center of Materials Science and Optoelectronics Engineering, University of Chinese Academy of Sciences,Beijing 100049, China}

\author{Ailun Yi}
\affiliation{State Key Laboratory of Materials for Integrated Circuits, Shanghai Institute of Microsystem and Information Technology, Chinese Academy of Sciences, 865 Changning Road, Shanghai 200050, China.}
\affiliation{Center of Materials Science and Optoelectronics Engineering, University of Chinese Academy of Sciences,Beijing 100049, China}

\author{Jiaxiang Zhang}
\affiliation{State Key Laboratory of Materials for Integrated Circuits, Shanghai Institute of Microsystem and Information Technology, Chinese Academy of Sciences, 865 Changning Road, Shanghai 200050, China.}
\affiliation{Center of Materials Science and Optoelectronics Engineering, University of Chinese Academy of Sciences,Beijing 100049, China}

\author{Chengli Wang}
\affiliation{State Key Laboratory of Materials for Integrated Circuits, Shanghai Institute of Microsystem and Information Technology, Chinese Academy of Sciences, 865 Changning Road, Shanghai 200050, China.}
\affiliation{Center of Materials Science and Optoelectronics Engineering, University of Chinese Academy of Sciences,Beijing 100049, China}

\author{Chun-Hua Dong}
\email[]{chunhua@ustc.edu.cn}
\affiliation{CAS Key Laboratory of Quantum Information, University of Science and Technology of China, 230026 Hefei, China.}
\affiliation{CAS Center for Excellence in Quantum Information and Quantum Physics, University of Science and Technology of China, 230026 Hefei, China.}

\author{Xin Ou}
\email[]{ouxin@mail.sim.ac.cn}
\affiliation{State Key Laboratory of Materials for Integrated Circuits, Shanghai Institute of Microsystem and Information Technology, Chinese Academy of Sciences, 865 Changning Road, Shanghai 200050, China.}
\affiliation{Center of Materials Science and Optoelectronics Engineering, University of Chinese Academy of Sciences,Beijing 100049, China}

\maketitle
\newpage
\noindent{\Large\bf{Supplementary Note S1. Splitting measurement}}
~\\

\par
\begin{figure*}[htbp]
	\centering
	\includegraphics[scale = 1]{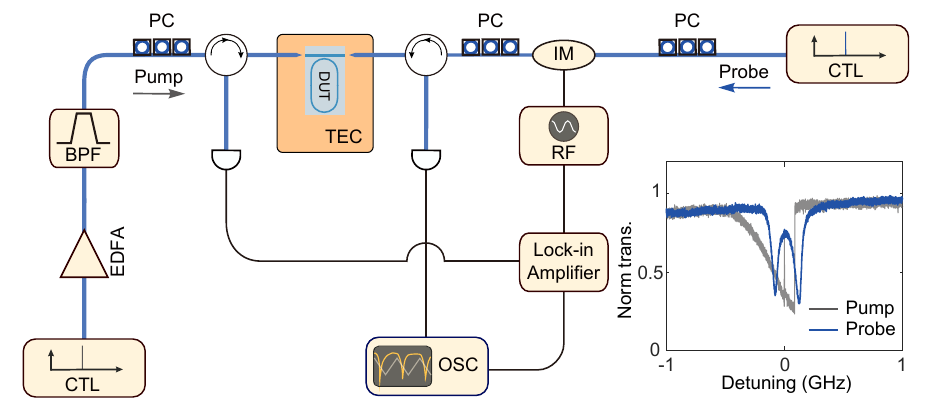}
        
	\caption{
        Modulator/Demodulator (MoDem) measurement for resonance splitting induced by photorefractive grating. TEC, thermoelectric cooler, DUT, device under test, CTL, continuously tunable laser, EDFA, erbium-doped fiber amplifier, BPF, bandpass filter, PD , photodetector, IM, intensity modulator, OSC, oscilloscope. The inset depicts the pump/probe transmission from two directions of the chip.
	}
	\label{figS2}
\end{figure*}

\large{The resonance splitting, which is directly due to the non-degenerate phenomena of two counter-propagating modes induced by cavity-enhanced photorefractive grating, can be quantified by literally pumping the racetrack resonators. However, as represented in \fref{figS2}, measured pump transmission features a large thermal-induced resonance distortion that influences the extraction of peak splitting value, thus the actual resonance shape should be detected by another laser with reduced optical power from the opposite edge facet. The experimental setup (\fref{figS2}) is similar to the previous work \textit{[Laser \& Photonics Reviews (2024): 2301351.]}. Using piezo-driven module, the sweeping pump laser activates the photorefractive grating effect in a \LT~microresonator after the EDFA, BPF and PC. To extract the mode splitting information, the probe laser that propagates in the backward direction is modulated by an electro-optic intensity modulator (IM). After injecting into the microresonator, the probe laser is then recieved by PD, and the probe signal mixed with strong backscattering pump light is recognized by the lock-in amplifier (Stanford Research Systems, SR844) based on the same applied RF signal.}

\noindent{\Large\bf{Supplementary Note S2. Loss and dispersion characteristics of \LT~microresonators}}
~\\
\par
\begin{figure*}[htbp]
	\centering
	\includegraphics[scale = 1]{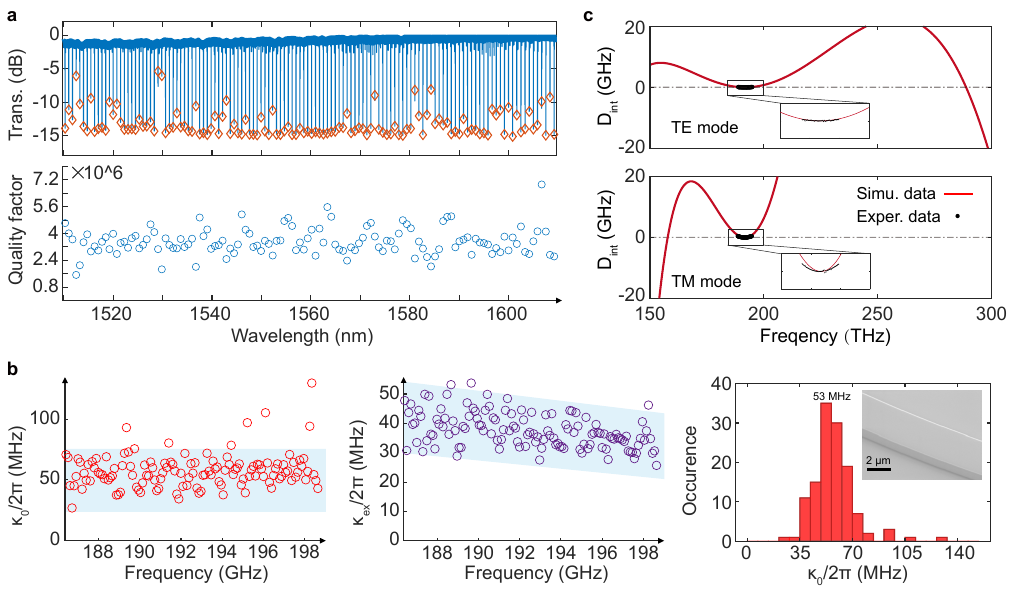}
	\caption{(a) Transmission of 100 GHz \LT~racetrack resonator with marks to resonance dips and Q-factors of corresponding resonances. (b) Fitted value of intrinsic linewidth $\kappa_{0}$ and external linewidth $\kappa_{ex}$, the $\kappa_{0}$ histogram indicates the most propable value of 53 MHz, corresponding to an approximate Q-factor of 3.6 million. (c) Simulated and measured dispersion data of TE modes and TM modes for efficient soliton generation.
	}
	\label{figS1}
\end{figure*}

By using a fiber polarization controller and lens fibers, a tunable TE-polarized laser is coupled in the \LT~resonators (2 $\mu$m multimode waveguide with 400 nm etched part and 200 nm slab part) to map the transmission spectra from 1510 nm to 1610 nm (\fref{figS1}(a)), which shows the desirable coupling. To further illustrate the cavity loss and its coupling state, the resonance linewidth compromising of external loss ($\kappa_{ex}$) and intrinsic loss($\kappa_{0}$) can be derived by:
\begin{equation}
T_{norm} = \frac{(\kappa_{ex} - \kappa_{0})^2 + 4\delta^2}{(\kappa_{ex} + \kappa_{0})^2 + 4\delta^2},
\label{eq1}
\end{equation}
where $\delta$ is the offset between resonance dip and laser frequency. As illustrated in \fref{figS1}(b), fitted results reveal its under-coupling feature that the declined $\kappa_{ex}$ with increasing frequency is lower than the $\kappa_{0}$. The constant range of $\kappa_{0}/2\pi$ value exhibits a probable average value of 53 MHz (the left and right panels in \fref{figS1}(b)), betraying near 3.6 million Q-factor of our tested \LT~cavity. The TE-mode and TM-mode dispersions are both simulated and measured in \fref{figS1}(c) to guarantee the proper integrated dispersion ($D_{int}$) for bright soliton seeding and to ensure the stronger anolomous dispersion for the prevention of the crosstalk associated with TM-mode microcomb generation, respectively, based on the equation \textit{[Nanophotonics 9.5 (2020): 1087-1104.]}: 
\begin{equation}
\omega_{\mu} = \omega_0 + \sum_{n = 1}^{\infty} D_{n}\mu^n = \omega_0 + D_1\mu + D_{int},
\label{eq1}
\end{equation}
where $\omega_0$ is the pump angular frequency, $D_1/2\pi$ is the cavity FSR and $\mu$ is the mode number.

\noindent{\Large\bf{Supplementary Note S3. Kerr soliton crystallization in X-cut \LT~microresonators}}
~\\
\par
\begin{figure*}[htbp]
	\centering
	\includegraphics[scale = 1]{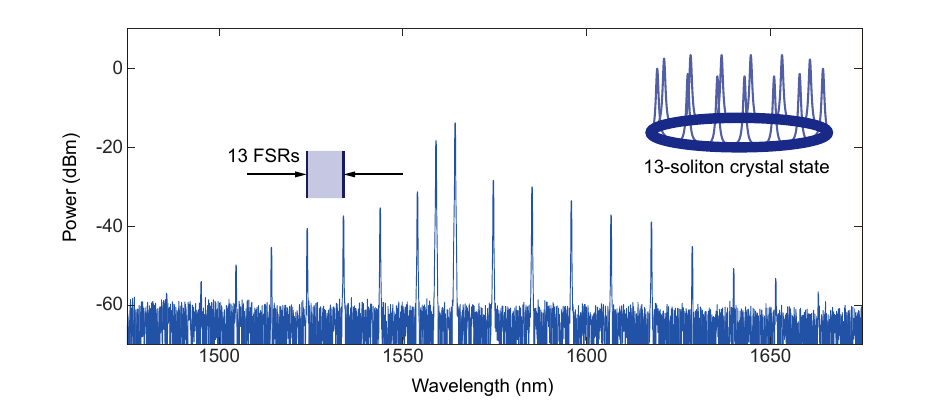}
	\caption{Optical spectrum of the 13-soliton crystal in X-cut \LT~platform. Inset: diagram of power waveforms in fast-time domain, presenting a soliton crystal with 13 distributed solitons.
	}
	\label{figS3}
\end{figure*}

Based on the dual-suppression strategy in this work (250 mW pump laser at 1564.23 nm and 400 mW auxiliary laser at 1559.178 nm), we obtain the generation of the soliton crystallization. As shown in \fref{figS3}, the frequency comb with 13 FSRs features a smooth sech-shaped spectral envelope that confirms the mode-locked soliton crystal. This crystallization, dominated by the interaction between the soliton formation and the avoided mode crossing, can be visualized to the periodic pulse train inside the resonators \textit{[Nature communications 12.1 (2021): 3179.,Nature Photonics 11.10 (2017): 671-676.]} (right inset in \fref{figS3}). 

Interestingly, the observation of crystallized soliton state should be the first time for the X-cut ferroelectric platform, and high pockels coefficient also empowers the chip-scale integration of manipulated multi-soliton molecules via electro-optic modulation \textit{[Optica 8.10 (2021): 1334-1339.,Optica 9.2 (2022): 240-250.]}.

\par

    
